# JMS: A workflow management system and web-based cluster front-end for the Torque resource manager

David K. Brown, Thommas M. Musyoka, David L. Penkler and Özlem Tastan Bishop*

Research Unit in Bioinformatics (RUBi), Department of Biochemistry & Microbiology, Rhodes University, Grahamstown, South Africa

**ABSTRACT**

**Motivation:** Complex computational pipelines are becoming a staple of modern scientific research. Often these pipelines are resource intensive and require days of computing time. In such cases, it makes sense to run them over distributed computer clusters where they can take advantage of the aggregated resources of many powerful computers. In addition to this, researchers often want to integrate their workflows into their own web servers. In these cases, software is needed to manage the submission of jobs from the web interface to the cluster and then return the results once the job has finished executing.

**Results:** We have developed the Job Management System (JMS), a workflow management system and interface for the Torque resource manager. The JMS provides users with a user-friendly interface for creating complex workflows with multiple stages. It integrates this workflow functionality with Torque, a tool that is used to control and manage batch jobs on distributed computing clusters. The JMS can be used by researchers to build and run complex computational pipelines and provides functionality to include these pipelines in external interfaces. The JMS is currently being used to house a number of structural bioinformatics pipelines at the Research Unit in Bioinformatics (RUBi) at Rhodes University.

**Availability:** The JMS is an open-source project and is freely available at https://github.com/RUBi-ZA/JMS

**Contact:** o.tastanbishop@ru.ac.za

## 1 INTRODUCTION

Computational pipelines or workflows have become an important tool for the analysis of the vast amounts of data being generated in the sciences. The computational complexity of these workflows varies widely, but can often require days of computing time and large amounts of computing power. To speed up the execution of these jobs, the use of parallel algorithms and computer clusters has become increasingly common. Computer clusters offer high performance through the aggregation of resources from multiple individual computers. Software is required to manage the submission and scheduling of jobs on these clusters as well as the allocation of resources to individual jobs. One such software system is Torque Resource Manager. Torque is open source software that provides fine-grained control over the resources of a cluster, allowing users to configure and manage nodes, submit jobs, and administer their systems. It can also be integrated with $3^{rd}$ party job schedulers such as Maui (Jackson et al. 2001) for improved job scheduling capabilities. Unfortunately, Torque presents quite a challenge to set up, requiring software to be manually installed and configured on each node of the cluster. In addition to this, Torque is made up of a number of different command-line tools, presenting a steep learning curve for the typical user.

In the biological sciences, a number of workflow management systems (WMS) have been developed over the past decade. Two such systems, Galaxy (Goecks et al. 2010) and Ergatis (Orvis et al. 2010), allow users to create workflows by piecing together a number of tools that come prepackaged with the systems**.** Users are also able to add additional tools by creating configuration files describing how these tools can be run. Unfortunately, these configuration files can be difficult for new users to master.

Workflow management systems like Ergatis and Galaxy manage the execution of tools and scripts directly on the servers or clusters where they are set up. There is another class of WMSs, however, that create pipelines out of web services. One example of this type of WMS is Taverna (Wolstencroft et al. 2013). Taverna takes advantage of the thousands of bioinformatics web services that have been developed over the past few years (Bhagat et al. 2010) by combining them, along with local scripts, into complex pipelines. The advantage of such an approach is that most of the computation is done on the remote servers where the web services are hosted, thus, reducing local infrastructure and maintenance costs. Limitations of this approach include bandwidth constraints as well as varying reliability of the remote web services.

---

*To whom correspondence should be addressed.

Although the afore mentioned WMSs do an admirable job of allowing users to build and manage workflows, they do not provide a means for monitoring and managing cluster resources at a fine-grained level. In addition to this, researchers who develop workflows often want to make these workflows available via their own web servers. In order to do this, they must spend time integrating the tool or workflow into their web server and developing an interface to it. In this paper, we introduce the Job Management System (JMS). The JMS combines the functionality of a cluster front-end with that of a WMS. It exposes this functionality via a RESTful web interface, allowing developers to include workflows that have been created within the JMS in their own web servers. The JMS also provides the ability to add new tools and scripts via the web interface, without any need for complicated configuration files.

Although the functionality of the JMS is applicable to any scientific field, it's currently being tailored towards bioinformatics with the introduction of bioinformatics related tools, workflows and result editors. As part of the H3ABioNet, a subset of the H3Africa Consortium (The H3Africa Consortium, 2014), the system will be open sourced and made available to groups around Africa, as well as the rest of the world, to ease the burden of setting up and managing new clusters. The JMS will further provide a work environment to share the tools and scripts between collaborating groups.

## 2 SYSTEM DESIGN & FEATURES

The JMS has been developed as a Django web application and provides an interface to a Torque-managed cluster (Fig. 1). It consists of two API modules, which provide all the functionality, and a third module that provides a web interface. A MySQL database is used to store workflows and a comprehensive job history for each user. A background service has also been developed that keeps the systems job history up-to-date.

### 2.1 User module

The first of the two API modules is the *user* module. This module is responsible for user authentication and security. It includes a custom-made authentication backend that allows users to authenticate against the Linux file system on the master node of the cluster. This allows the JMS to assume the user's identity when running processes on the server or submitting jobs via Torque. By using the Linux permissions system, users are able to access, via the JMS, what they would be able to access if logged into the server via SSH, for example.

The *user* module also offers some social networking features, allowing users to communicate with each other, create groups, and share files. It can be accessed via a RESTful web API.

### 2.2 Job module

The *job* module contains all the functionality required to create and manage workflows, submit jobs, and interface with Torque. Like the *user* module, the *job* module can be accessed via a RESTful web API.

#### 2.2.1 Dashboard

The JMS provides a dashboard for monitoring the status of jobs. Amongst other things, the dashboard provides users with summary information for the status of the cluster as a whole. This information includes how many nodes are online/offline, the proportion of processors being used across the whole cluster, the number of jobs currently running or waiting to run, and the amount of disk space still available.

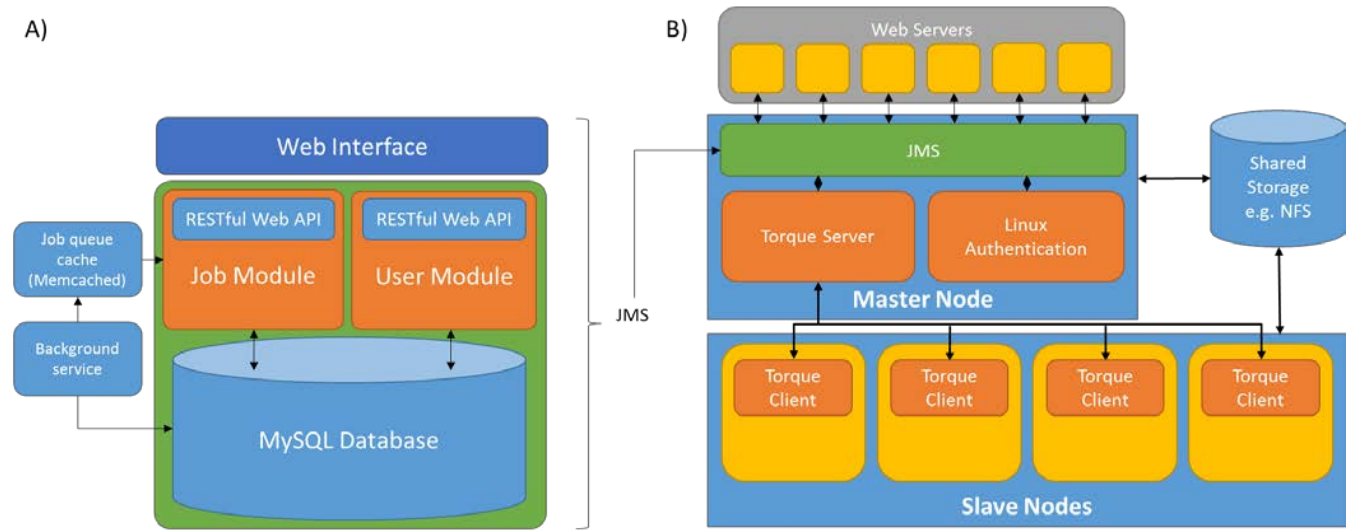

**Fig. 1. A)** The JMS has been developed as a Django web application. The Django project consists of 3 modules and a MySQL database. The j*ob* module is the main module and is responsible for interfacing with Torque as well as providing the WMS functionality. For improved performance, the *job* module interacts with an in-memory cache instead of the database for certain transactions. A background service is used to update job history data in both the cache and the database. The *user* module is responsible for handling user authentication and security. It also provides basic social networking functions. Both these modules expose their functionality via a RESTful web API. The last module makes use of these web APIs to provide a web-based GUI for the system. **B)** The JMS provides an interface for external web servers to run jobs on a cluster through the Torque resource management software. Authentication is done via the Linux file system so that users have the same permissions they would have if they logged into the server via SSH.

The dashboard also displays summary information for each node of the cluster. This includes how many processing cores the node has available and what particular jobs are running on the node.

Lastly, the dashboard also provides the current queue of jobs submitted to the cluster. This information includes the job ID, job name, the owner of the job, and the resources being used by the job.

*2.2.2 Job Management*

The JMS allows users to submit new jobs to the cluster, monitor and manage jobs while they run, and obtain the results of the jobs once they are complete. It does this by interfacing with various Torque commands, as well as the workflow management system, which will be discussed in section 2.2.3.

The JMS allows users to upload or create scripts to be submitted to the cluster and then request resources such as required memory, number of cores, and the wall-time to be allocated to the job. Based on these inputs, the JMS generates a job script and submits it to the cluster. Jobs can then be monitored until their completion. The JMS monitors the resources used, the input and output streams of the job, and the working directory of the job. On completion, the user can access the results of a job, either from the output and error streams, or by downloading the resultant files from the working directory.

The monitoring of jobs run on the cluster is done by interfacing with the `qstat -f` command. This command is run every time a job starts or ends via prologue and epilogue scripts. Data returned is parsed and stored in the JMS database in order to keep a permanent record of all jobs run. In addition to running this command at the beginning and end of each job, a background service runs the command at scheduled intervals (every 30s by default), continuously updating the records in the database while jobs run. Each time a user requests an update on the job, the data is then obtained from the database instead of running the `qstat -f` command. If a large number of users are monitoring jobs at the same time, it may create a substantial number of unnecessary hits on the database. To reduce database traffic, the JMS uses Memcached (Fitzpatrick 2004) to create an in-memory cache of the job history.

Prologue and epilogue scripts are a feature of Torque. They are scripts that run at the beginning and end of a job and are often used to clean up the working directory or record some accounting information. The JMS sets up prologue and epilogue scripts that allow it to monitor all jobs that are submitted to the cluster, even if those jobs are not submitted via the JMS interface.

In addition to monitoring jobs, the JMS also allows users to manage their jobs. Users may cancel running jobs, suspend or hold jobs before or after they start running, and, in future, users will be able to make requests to alter jobs. Because alteration requests may include requests for additional resources etc., they require an admin user's approval. If a non-admin user makes a request for an alteration to one of their jobs, the request will be forwarded to an admin user to be granted or denied.

*2.2.3 Workflow Management*

In addition to interfacing with Torque, the JMS provides additional functionality allowing users to build and execute complex computational pipelines or workflows.

In the JMS model of a workflow, a workflow is made up of a set of stages. A stage can be a command-line utility that is already installed on the cluster or a custom script or executable that is uploaded by the user. For each stage of a workflow, users provide the JMS with a number of details including the command that would be used to run the job from the terminal, the parameters that the command can take, the resources that should be allocated to each stage in the workflow, and which stages in the workflow a particular stage depends on. Optionally, users can also specify files they expect to be generated as outputs.

The JMS makes it easy for users to develop complex workflows by creating conditional stages. This is done by setting conditions on stage dependencies. For example, given 3 stages, A, B, and C, a user can set B to run if A completes successfully and C to run if A fails (Fig. 2. A). This functionality comes from Torque, but the JMS extends this further by allowing users to create conditions based on the exit status of a stage. For example, B executes if the exit status of the job is 1 and C executes if the exit status of the job is 2 (Fig. 2. B). Users can take advantage of this functionality by manually setting the exit status of their scripts based on some condition.

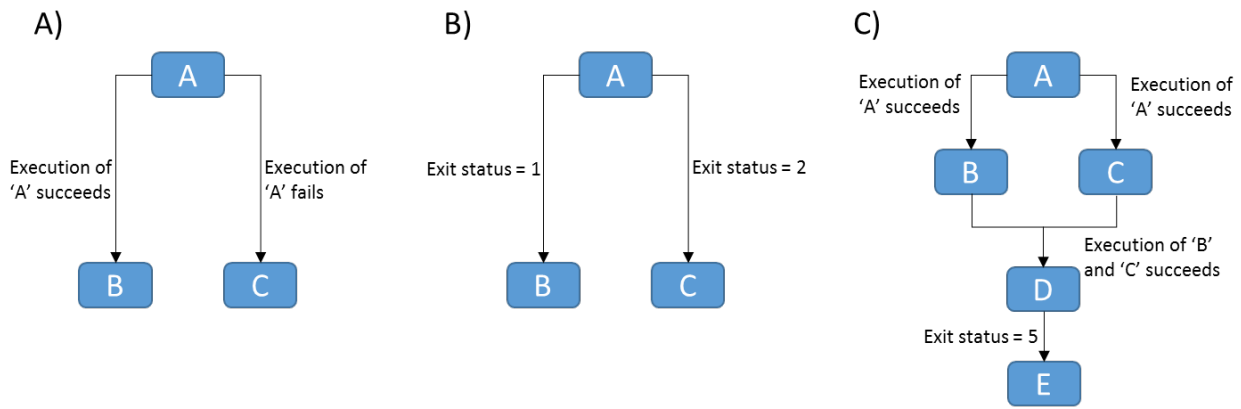

**Fig. 2. A)** Stage B is executed if stage A executes successfully, else stage C executes. **B)** Stage B will only execute if stage A exits with a status code of 1 and stage C will only exit if stage A exits with a status code of 2. If the status code is neither 1 nor 2, the job fails. **C)** All other stages wait while stage A executes. On successful completion of A, stage B and C execute in parallel. If both stages execute successfully, stage D executes. If stage D exits with a status code of 5, stage E executes.

The JMS also allows for certain stages in a workflow to run in parallel while others are required to run sequentially. For example, if we have stages A to E, we could have a scenario where B and C are dependent on A executing successfully, D is dependent on B and C executing successfully, and E is dependent on D exiting with a status code of 5 (Fig. 2C). In such a case, all stages would wait while A executes. On successful completion of A, B and C would both execute and run in parallel. On successful completion of both B and C, D would begin execution. If D exits with a status code of 5, stage E will begin executing. In this example we see that certain stages must execute within sequence, while others (B and C) may execute in parallel.

The web-based interface to the JMS (discussed further in section 2.3) provides a user-friendly, click-based interface for creating workflows. Unlike WMSs like Galaxy and Ergatis, the JMS does not require users to deal with configuration files. All workflow data is stored in the JMS database. Scripts and executables that are uploaded by users are automatically stored in a strict directory hierarchy that is managed by the JMS. In addition to uploading scripts, users can also create new scripts from within their browsers. Scripts can also be edited via the web interface. This allows users to update their scripts or fix bugs without needing to re-upload the script.

Once a user has created a workflow, it can be shared with other users. The creator of the workflow can control what permissions these users have with regards to the shared workflow. Users can be given permission to edit the workflow as well as run it, or they can be limited to simply being able to view it or not have access to it at all.

To run a workflow, the JMS provides an automatically generated web interface. This interface allows users to enter in values for each of the parameters specified during the creation of the workflow. Although other WMS also provide automatically generated interfaces, the JMS allows these interfaces to be integrated into other web servers (see section 4.3 for an example). A JavaScript plugin is currently under development that will automate the process of generating an interface for a 3[rd] party website.

Some workflows may consist of a number of different stages, each of which requires a number of different parameter values to be input. To make the process of inputting parameters faster and easier, as well as more consistent, the JMS allows users to create input profiles. An input profile is basically a set of default inputs for a workflow. Users can create multiple input profiles for each workflow. This is especially useful when a user wants to run a workflow multiple times, where each time only one or two parameters need to be changed. An input profile can be created that automatically fills in values for the parameters that are to remain constant, and the user only needs to fill in values for the parameters that will change.

Although input profiles will significantly speed up the rate at which multiple jobs can be submitted, this may still not be good enough for cases when hundreds or thousands of jobs need to be submitted. The JMS caters for these cases by allowing users to submit batch jobs. Batch jobs require the user to generate a batch file and submit it to the JMS. The JMS reads in the file and submits each job, one at a time.

The job history stored for workflows is also enriched by the JMS. In addition to the cluster-related information obtained from Torque, all the data for each stage of the workflow run is stored. This includes all the parameter values that were entered by the user, as well as a snapshot of the working directory after each stage. This means jobs can be repeated from any stage.

The JMS also provides functionality for workflows to be exported to a compressed file and downloaded. Users can then keep this file as a backup or distribute it to other users who can import it into their JMS instances.

### 2.2.4 Cluster Configuration

In addition to interfacing with Torque to provide job management functionality, the JMS provides cluster configuration functionality. This allows admin users set up and manage queues and configure server settings. Adding and configuring a new node using the Torque command-line clients can be a difficult task for a new user. All the correct Torque software and configuration files have to be set up on the new node before the master node will recognize it. The JMS eases the burden of adding new nodes to the cluster by automating certain parts of the process and then providing manual instructions to administrators to finish of the process.

### 2.3 The interface

The final Django module is the *interface* module. This module provides a comprehensive web interface that interacts with the other two modules by using AJAX to access their RESTful web APIs. As such, all the functionality provided by the modules described above is also available via the web interface. This includes submitting, managing, and monitoring jobs, creating and managing workflows, and cluster configuration and monitoring.

The *interface* is predominantly a client-side module and contains very little server-side code. The Knockout.js JavaScript framework has been employed on the client-side to make the JavaScript code more concise and manageable.

By exposing a web API, the *job* and *user* modules also allow 3[rd] party developers to develop their own interfaces for the system. This will also allow us to build additional interfaces for other platforms (e.g. mobile platforms) in the future.

### 3 USE OF THE JMS

The JMS is an open-source project and can be downloaded from https://github.com/RUBi-ZA/JMS. Information on how to install the system can also be obtained from here.

The JMS is distributed with Torque v5. The JMS install script can optionally install Torque as well. On installation, a strict directory hierarchy is created for the purpose of storing user-created workflows as well as job files. Once installed, the Django web application will need to be hosted using a web server of choice (testing has been done using Apache2). An example Apache host configuration file is shipped with the project. Once this is set up, users can access the JMS by browsing to the relevant URL.

Because the JMS authenticates against the Linux file system, user accounts are created from within Linux. This means that anyone who has an account on the machine, automatically has access to the JMS.

After logging in, users are redirected to the Dashboard where they can monitor their jobs in the queue as well as the status of the cluster. They are also able to cancel their jobs from this page.

Selecting a job in the queue will take the user to the Job History page where more detailed information about the job can be found. More options for interacting with the job can be found here such as deleting, repeating and sharing jobs.

From the Workflows page, users can create new workflows, edit and run existing workflows, or delete unwanted workflows. Users can also import and export workflows from this page as well as share workflows with other users.

The final page on the site is the Settings page. This is where users can perform cluster configuration, add nodes, and set up and manage queues. This page is only available to admin users.

Developers can integrate workflows into their own websites by creating their own interface that interacts with the JMS web services via AJAX. A JavaScript plugin is currently under development that will automatically generate an interface for a given workflow.

## 4 WORKFLOWS

Below we provide two structural bioinformatics pipelines as examples of how the JMS is currently being used within our group. However, the JMS is not limited to running purely scientific workflows, but also can be used as a workflow manager system. An example currently under development is a database of natural compounds. Once it is public, users will be allowed to upload compounds to the database via the JMS system.

### 4.1 Protein-ligand docking

A pipeline for *in silico* docking by AutoDock4 (Morris et al. 2009) has been developed using the JMS. The workflow accepts an *apo* protein receptor and ligand as separate PDB files, and by utilizing several AutoDock4 scripts, automates a "blind docking" experiment (Fig. 3A). The ligand in such an experiment is exposed to the entire surface of the protein. To achieve this, a grid box must be generated to encapsulate the entire protein, and must therefore be centered on the protein's center of mass. Thus in this workflow, the grid center is automatically set to 'auto'. Given that the size of different protein receptors may vary considerably from one to another, the size of the grid box is defined by the user and set via the GUI interface by entering in the number of points and the required spacing. Should the user want to specify their own docking parameters, they can be uploaded in .txt file format as is customary with AutoDock4. Once these requirements are met, the entire docking process is completely automated, from the preparation of the receptor protein and ligand, to setting up the grid parameters and generating the grid box, and finally preparing the docking parame-ters and running AutoDock4. On completion, a docking log file is written that contains all the docking results.

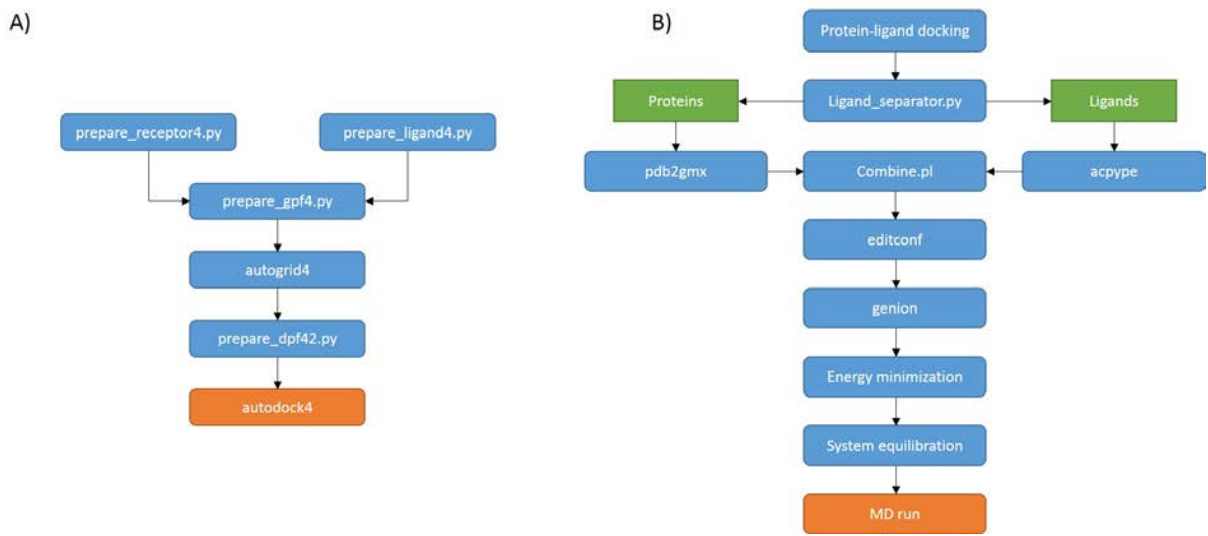

**Fig. 3. A)** The docking pipeline is made up strictly of AutoDock4 scripts and executables. It automates the entire docking process from ligand and receptor preparation to the actual docking runs. **B)** GROMACS is used to perform molecular dynamics simulations on the results from protein-ligand docking

## 4.2 Molecular dynamics

The JMS is used to house a pipeline that utilizes GROMACS 4.5 software (Pronk et al. 2013) as well as the *g_mmpbsa* tool (Kumari et al. 2014) to perform molecular dynamics on a protein-ligand complex (Fig. 3B). This pipeline can be used in conjunction with the docking pipeline described above. After docking, the protein-ligand complex is split into separate coordinate files. This is necessary as the ligands atoms have to be parameterized using an external tool - ACPYPE (Wang et al. 2004; Wang et al. 2006) - before GROMACS will recognize them. The protein topology files can be generated by the *pdb2gmx* tool provided by GROMACS. Using a python script, the protein generated .gro file is combined with that of the ligand and solvated in a box (type and size specified by user depending on the size of the protein) using the *editconf* tool. The system is neutralized by adding counter ions ($Na^+$ or $Cl^-$) via the *genion* tool. The system energy is subsequently minimized up to a user defined tolerance and equilibrated via the NVT and NPT canonical ensemble. The equilibrated system is then subjected to a production run whose time length is defined by user. The final MD trajectory is analyzed using GROMACS tools that have been automated using Perl and Python scripts.

## 5 DISCUSSION

The JMS provides three distinct categories of functionality. Firstly, it provides a web interface to the Torque resource manager with the aim of making Torque much easier to work with. The few existing, free, web-based interfaces to Torque provide only basic functionality. The JMS interfaces with a substantial number of the Torque command-line clients, providing a number of job management and cluster configuration capabilities.

Secondly, The JMS provides a fully functional workflow management system, which integrates directly with Torque, further extending the abilities of the resource manager. The WMS provides advanced scheduling and dependency abilities, input profiles, batch jobs, and, we believe, a far easier way of building and managing workflows than existing systems provide.

Lastly, the JMS allows workflows to be integrated into other, external websites via its RESTful web API. A JavaScript plugin is being built to automatically generate interfaces for these websites.

Using the JMS, researchers are able to create workflows via the click-based web interface and then easily integrate these workflows into their own sites. This functionality substantially reduces the effort of creating web interfaces for computational pipelines. Further, we believe that the JMS will facilitate the sharing of information and computational techniques within H3Africa collaborations and help with the setting up of cluster related infrastructure.

The inclusion of Torque administration features means that the JMS is useful to both casual users, who simply want to run jobs over their cluster, and system administrators, who can use it for monitoring and managing the cluster.

Currently, the JMS must be installed on the master node of the cluster and can be used to manage Torque servers located on that machine. Future work will extend the JMS to allow it to manage multiple clusters in remote locations. In line with this, we will also provide advanced scheduling functionality that will allow users to schedule workflows across multiple clusters.

In addition to this, we will develop JMS apps for smartphones that will allow normal users to monitor and submit jobs from their phones and admin users to receive status notifications and requests for resources.

Lastly, because we store snapshots of jobs at each stage, a single job can end up taking a lot of storage space. In future, the JMS will allow admin users to set up an automatic backup and archiving solution, whereby old jobs are compressed and copied to an external location.

## 6 ACKNOWLEDGEMENTS

We would like to thank all the members of the Research Unit in Bioinformatics (RUBi) at Rhodes University for their valuable time and advice.

*Funding*: *This work is supported by the National Institutes of Health Common Fund under grant number U41HG006941 to H3ABioNet. The content of this publication is solely the responsibility of the authors and does not necessarily represent the official views of the National Institutes of Health.*

*Conflict of Interest*: None declared.

## REFERENCES

Bhagat, J. et al. (2010) BioCatalogue: a universal catalogue of web services for the life sciences. *Nucleic Acids Research*, 38, W689–94.

Fitzpatrick, B. (2004) Distributed caching with memcached. *Linux Journal*, 2004, 5.

Goecks, J., Nekrutenko, A. and Taylor, J. (2010) Galaxy: a comprehensive approach for supporting accessible, reproducible, and transparent computational research in the life sciences. *Genome Biology*, 11, R86.

Jackson, D., Snell, Q. and Clement, M. (2001) Core Algorithms of the Maui Scheduler. *Job Scheduling Strategies for Parallel Processing*, 2221, 87–102.

Kumari, R., Kumar, R. & Lynn, A. (2014) G-mmpbsa -A GROMACS tool for high-throughput MM-PBSA calculations. *Journal of Chemical Information and Modeling*, 54, 1951–1962.

Morris, G. M. et al. (2009) AutoDock4 and AutoDockTools4: automated docking with selective receptor flexiblity. Journal of Computational Chemistry 2009, 16, 2785-2791. Orvis, J. et al. (2010) Ergatis: A web interface and scalable software system for bioinformatics workflows. *Bioinformatics*, 26, 1488–1492.

Pronk, S. et al. (2013) GROMACS 4.5: A high-throughput and highly parallel open source molecular simulation toolkit. *Bioinformatics*, 29, 845–854.

The H3Africa Consortium (2014) Research capacity. Enabling the genomic revolution in Africa. *Science (New York, N.Y.)*, 344(6190), 1346–8.

Wang, J. et al. (2006) Automatic atom type and bond type perception in molecular mechanical calculations. *Journal of Molecular Graphics and Modelling*, 25, pp.247–260.

Wang, J. et al. (2004) Development and testing of a general Amber force field. *Journal of Computational Chemistry*, 25, 1157–1174.

Wolstencroft, K. et al. (2013) The Taverna workflow suite: designing and executing workflows of Web Services on the desktop, web or in the cloud. *Nucleic Acids Research*, 41, (W1): W557-W561.